\documentclass{epl}

\usepackage{amsmath,amssymb}

\usepackage{graphicx}

\newcommand{\bra}[1]{\left<#1\right|}
\newcommand{\ket}[1]{\left|#1\right>}

\newcommand{\br}{{\bf r}}
\newcommand{\bk}{{\bf k}}

\newcommand{\be}{{\bf e}}

\newcommand{\bA}{{\bf A}}

\newcommand{\f}{\frac}

\newcommand{\df}{\nabla}

\newcommand{\Eq}[1]{Eq.~(\ref{#1})}
\newcommand{\Fig}[1]{Fig.~\ref{#1}}
\newcommand{\Ref}[1]{(\ref{#1})}

\newcommand{\av}[1]{\left<{#1}\right>}

\newcommand{\pref}[1]{(\ref{#1})}

\newcommand{\ee}{\end{eqnarray}}
\newcommand{\bee}[1]{\begin{eqnarray}\label{#1}}

\begin{document}

\title{Topological Order in the (2+1)D
       Compact Lattice Superconductor}

\shorttitle{Topological order...}

\author{Anders Vestergren\inst{1} \and Jack Lidmar\inst{1}
  \and T. H. Hansson\inst{2}}


\institute{
\inst{1}
Department of Physics,
Royal Institute of Technology,
AlbaNova, 
SE-106 91 Stockholm,
Sweden\\
\inst{2}
Fysikum,
Stockholm University,
AlbaNova, 
SE-106 91 Stockholm,
Sweden
}
\pacs{74.20.-z}{Theories and models of superconducting state}
\pacs{11.15.Ha}{Lattice gauge theory}
\pacs{71.10.Hf} {Non-Fermi-liquid ground states, electron phase diagrams
	 and phase transitions in model systems}

\date{\today}
\maketitle
\begin{abstract}
We study topological aspects of a compact lattice superconductor, and
show that the characteristic energy splitting, $\Delta$, between
almost degenerate ground states, is simply related to a novel order
parameter $\tilde W$, which is closely related to large Wilson loops.
Using Monte Carlo methods, we study the scaling properties of $\tilde
W$ close to the deconfining phase transition, and conclude that
$\Delta \sim e^{-L/\xi}$, where $L$ is the size of the system, thus
giving quantitative support to the vortex tunneling scenario proposed
by Wen.
\end{abstract}

The concept of topological order was first introduced to describe the
ground states of the quantum Hall system~\cite{wenrev}.  Although the
present interest in topological order to large extent derives from the
quest for exotic non-Fermi liquid states of relevance for the
high-$T_{c}$
superconductors~\cite{nagaosa:966,balents,senthil-fisher,herbut,sachdev03},
the concept is of much more general interest.  Both spin
liquids~\cite{wen-spinliq,moessner-sondhi}, believed to be of
relevance for frustrated magnets, and ordinary BCS superconductors
with dynamical electromagnetism~\cite{wen91,hos2004}, are examples of
topologically ordered systems.
Two characteristics of topological order are particularly striking;
 fractionally charged quasiparticles,  and a  
ground state degeneracy depending on the topology of the 
underlying space, which is  lifted by tunneling processes. 
In the case of the most celebrated example -- the
Laughlin FQH states -- both these properties are well understood~\cite{wen90}, 
but for  superconductors, the situation is less clear. It was only in 1990
that Kivelson and Rokshar pointed out that the quasiparticles are
indeed fractional~\cite{kiv90}, and to our knowledge there has been no
quantitative study of ground state degeneracy and splitting.

We improve on this situation by a numerical study of a 2D compact
lattice superconductor, using a novel order parameter $\tilde W$,
which is sensitive to vortex tunneling processes. Our results are in
{\em quantitative} agreement with predictions based on topological
order, and rules out the possibility of a spontaneously broken
discrete symmetry.  A compact lattice superconductor captures
important properties of real type II superconductors~\cite{hos2004},
and
the 2D case 
is also of intrinsic interest in the context of
effective gauge theories for the high-$T_c$
cuprates~\cite{balents,senthil-fisher,herbut,nagaosa:966,sachdev03}.
Examples of recent proposals 
are the $U(1)$ slave-boson theory~\cite{nagaosa:966}, the
nodal liquid and the spinon-chargeon $Z_2$
theory~\cite{balents,senthil-fisher}, and compact
QED$_3$\cite{herbut}.

We use the Villain form of the action~\cite{ein78},
\bee{act}				
  S =   \f{J}{2}\sum_{\br \mu}
 \left ( \df_\mu \theta_\br + 2\pi n_{\br \mu} - q A_{\br \mu}
  \right)^2 
   + \f{1}{2g}\sum_{\br \mu} \left( B_{\br \mu} + 2\pi k_{\br \mu}
  \right)^2 \,  ,
\ee
where the charge $q$ phase field $\theta_\br$ is defined on the sites
$\br=(x_0,x_1,x_2)$, and the compact vector potential $A_{\br \mu}$ on
the links $(\br ,\br+\be_\mu)$, of a Eucledian 3D cubic lattice.
$B_{\br \mu} = \epsilon_{\mu\nu\lambda}\df_\nu A_{\br \lambda}$ is the
dual field strength, which is defined on the links of the dual lattice
\footnote{We use a continuum notation whenever convenient, and the
lattice version should be obvious. }.

We shall in the following 
concentrate on
$q=2$ pertinent both to a Cooper paired BCS superconductor and to the
more exotic pairing scenarios in effective gauge
models~\cite{senthil-fisher,herbut,nagaosa:966,sachdev03},
although we will keep a general $q>1$ where appropriate.
The phase diagram of the
$q=2$ model is well established~\footnote{ The phase diagram in the
$q=1$ case is controversial.}:
In the strong coupling limit $J \rightarrow \infty$, \pref{act}
becomes a $Z_{2}$ gauge theory, which has a phase transition at finite
$g$.  This transition between the ``confined'' phase at large $g$ and
the superconducting, or Higgs, phase at small $g$, can rigorously be
shown to extend to finite $J$~\cite{frad79}, and is believed to
extend all the way to the 3D $XY$ transition in the $g\rightarrow 0$
limit.  This is confirmed by recent simulations~\cite{Sudbo}.
The line at $J=0$ corresponds to compact QED which is
confined at all $g$ because of instanton effects~\cite{poly77}.

The topological structure of the theory, which is our present concern,
is particularly simple in the $g \rightarrow 0 $ limit of the $Z_{2}$
line, where the ground state degeneracy on, say, a torus can be
understood in terms of ``visons'' describing $Z_{2}$ magnetic fluxes
through the ``holes''~\cite{senthil-fisher,wen-spinliq}.  At finite
$g$ the ground states are connected by processes where these vortices
tunnel around the closed cycles of the torus, leading to a predicted
energy splitting $\Delta\sim e^{-cL/\xi}$~\cite{wen91}.  Close to the
phase transition, the vortices begin to proliferate and this simple
formula might well break down.  Note that in the case of spontaneous
breakdown of a discrete symmetry, there would also be tunneling
processes connecting different ground states, but with a splitting
$\Delta\sim e^{-cL^2/\xi^2}$ characteristic of domain wall
tunneling~\cite{wen91}.

To move away from the limiting cases, and to study the region close to
the phase transition, which is of particular interest in the models of
high-$T_c$ superconductivity referred to above, numerical simulations
are mandatory.
It  will be advantageous to use a dual formulation in terms
of flux tubes and monopoles (\emph{i.e.}\ instantons)~\cite{ein78,pesk78}
and express the partition function for \pref{act}  as
$
Z = \sum_{\left\{ \mathbf{m}_{\br},\, N_{\br} \right\}} e^{-S}
$,
with 
\bee{duals}				
  S = 2\pi^2J \sum_{\br \br'}  \mathbf{m}_{\br} \cdot V_{\br \br'} \mathbf{m}_{\br '}
    + q^2 \lambda^2  N_{\br} V_{\br \br'} N_{\br '} \, .
\ee
Here $\mathbf{m}_{\br} \in \mathbb{Z}^3$ is the vorticity on the dual links
and $N_{\br} \in \mathbb{Z}$ is the number of monopoles on the dual
sites.  The flux lines are constrained to start or end on the
monopoles only in quanta of $q$, \emph{i.e.}, $\nabla \cdot \mathbf{m}_{\br} =
qN_{\br} $.  The interaction is given by $ V_{\br } = L^{-3} \sum_\bk
V_\bk e^{i \bk \cdot \br } $, with $ V^{-1}_\bk = \sum_\nu
4\sin^2(k_\nu /2) + \lambda^{-2} $, which for large distances is a
screened coulomb interaction $V(r) = e^{-r/\lambda}/4\pi r$ with a
screening length $\lambda$ given by $\lambda^{-2} = J g q^2$.  In the
limit $\lambda \to 0$ the interaction reduces to an on-site
interaction, 
\bee{stronga}
  S = \f{\Phi_0^2}{2g} \sum_\br  \mathbf{m}_\br^2 \, ,
\ee
where the flux quantum $\Phi_0 = \frac {2\pi} q$.
Since each $+$($-$) monopole has
precisely $q$ outgoing (incoming) flux lines, there are (for $q>1$)
two distinct phases: In the superconducting phase,
where flux lines 
cost a lot of energy,
the monopoles are confined in neutral
pairs bound together by $q$ flux lines.  In the other phase, the flux
lines
condense, connect many monopoles, and form a large connected tangle
which percolates through the whole system.  The electric properties of
these phases are encoded in the behavior of a large Wilson loop $W =
\av{\exp\left(i \oint_\mathcal{C} \bA \cdot d\br \right)}$, for
fractionally charged test particles.  In the percolating flux phase,
where large flux loops dominate, the flux on any patch the size of a
correlation length in a surface spanned by $\mathcal{C}$, can be
considered as a random variable, thus giving an area law, while in the
superconducting phase, the small loops will only contribute close to
the edge of $\mathcal{C}$, giving a perimeter law~\cite{ein78}.

Using periodic boundary conditions on the
$\mathbf{m}$:s,  the total flux for a configuration 
is just the sum of the  $\mathbf{m}$:s through a full cross-section
$\mathcal{S}$ of the system in the direction $\mu$, 
 $\Phi_0 M_\mu = \Phi_0 \sum_{\br
\in \mathcal{S}} m_{\br\mu}$. 
Clearly $\tilde W_\mu = \av {e^{i \Phi_0 M_\mu} }$ is
closely related to large Wilson loops and
directly measures the presence of percolating flux, 
so we expect it to have area law behaviour in 
the confined phase. In the superconducting phase, however,
$W$ and $\tilde W$ differ --  the latter 
is insensitive to small flux loops and
should not obey a perimeter law.
For these reasons, we submit that $\tilde W$ is a good (non-local) order
parameter for the deconfinement transition, and if the transition is
continuous, we expect a finite size scaling relation,
\begin{equation}					\label{W-scaling}
  \tilde{W}_\mu(\delta,L) = \av{e^{iM_\mu\Phi_0}} = \hat{W}_\pm(L/\xi),
\quad
  \xi \sim |\delta|^{-\nu},
\end{equation}
to hold.  Here $\delta$ is a tuning parameter of the transition, {\em
e.g.}, $\delta = (J - J_c)/J_c$, $\hat{W}_\pm$ is a universal scaling
function, and $\xi$ is the correlation length which diverges as the
transition is approached\footnote{
For $q=1$ this would  not give any information,
since in this case $\tilde W = 1$ identically. }.

\begin{figure}
\centerline{\includegraphics[width=7cm]{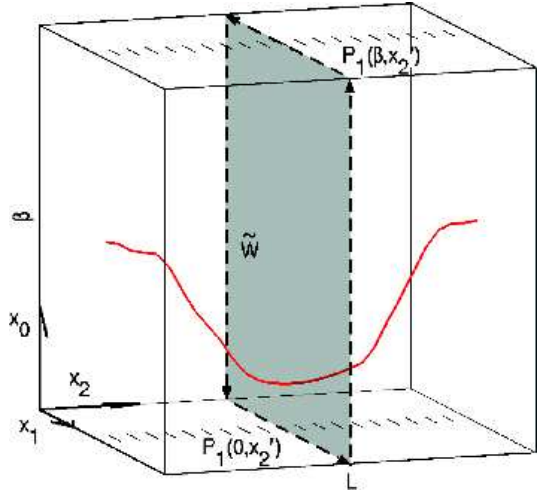}}
\caption{
\label{tunnel}
Eucledian space-time configuration corresponding to a vortex tunneling
around a closed cycle in the $x_2$ direction on the torus.  The
operator $\tilde W = \av{\exp(i\oint_\mathcal{C} \bA \cdot d\br)}$
measures the flux through the loop surrounding the shaded area in the
figure.  The loop may be decomposed into two vertical Polyakov loops,
$P_0$ and $P_0^\dagger$, which cancel each other for periodic boundary
conditions in the spatial directions, and two horizontal ones, $P_1$
and $P_1^\dagger$, that differ only on a single twisted link belonging
to the indicated column that is twisted by the gauge transformation
$\Omega_{10}$.
}
\end{figure}

Physically, configurations having non-zero total flux $\Phi_0 M_\mu$
around spatial directions of the torus correspond to vortex tunneling
events.  In order to include these in the partition function
we must be careful about boundary conditions.
Returning to the original formulation in terms of $A$ and $\theta$,
we note that periodic boundary conditions in the $\mu$ and $\nu$
directions are compatible only with the total vorticity in the $\mu\nu$ 
plane being an integer multiple of $q$. 
We can, however, choose twisted boundary conditions,
\bee{tbc}
A_{\br\mu}|_{_{ x_\nu=L} }= \Omega_{\mu\nu}(n_{\mu\nu}) 
A_{\br\mu}|_{_{ x_\nu=0} } \, ,
\ee
where $\Omega_{\mu\nu}$ is a large gauge transformation, that shifts
$A_\mu$ on all links in one particular column in the $\sigma$
direction, ($\mu\nu\sigma$ cyclic) by $2\pi n_{\mu\nu}/q$, with
$n_{\mu\nu} = 1,\dots, q$.  The twisted links corresponding to
$\Omega_{10}$ are shown in \Fig{tunnel}.  
Acting as an operator on states in the Hilbert space
$\Omega_{\mu\nu} $ changes the value of the Polyakov loop operator
$P_\mu =\exp({i\oint A_\mu})$ (the integration is around a
closed cycle on the torus), according to the topological algebra
$\Omega_{\nu\mu} P_\nu - e^{i 2\pi n_{\mu\nu}/q} P_\nu \Omega_{\nu\mu}
=0 $.  With the boundary conditions \pref{tbc}, the total vorticity in
the $\sigma$ direction is $n_{\mu\nu}+ n_{\nu\mu}\ ({\rm mod}\ q)$.

In order to discuss quantum mechanical states, we now distinguish
between the Euclidean time direction $\mu =0$, and the spatial
directions $i=1,2$.  For simplicity we also specialize to $q=2$ and put
$\Omega_{ij}=\Omega_{ij}(1)$.  Periodicity in space implies even total
flux in the $0$-direction, and, absent vortices, the Polyakov loop
operators, $P_i$ can be diagonalized, with eigenvalues $\pm 1$, so the
ground state manifold is spanned by the states $\ket{f_1,f_2}$, where
$\Omega_{10} \ket{1,1} = \ket{0,1}$ {\em etc}. Tunneling matrix
elements between these states can be extracted from the twisted
partition functions,
\bee{twist}
Z_{odd}^i = {\rm Tr}\left[ e^{-\beta H}\Omega_{i0}\right].
\ee
The corresponding Euclidean path integral with the boundary condition
\pref{tbc} in the time direction will, in the flux-monopole
formulation, amount to summing over configurations with odd total flux
through the $(0i)$ plane.  Similarly, the untwisted partition function,
$Z_{ev.}^i$, corresponds to even total flux.  \Fig{tunnel} illustrates
that $Z_{odd}^1$ indeed corresponds to vortex-antivortex pairs
tunneling in the $2$-direction.
In the presence of vortex loops and monopole-antimonopole pairs, 
the ground states are no longer eigenstates of the loop operators 
$P_i(x_0,x_j)$. 
Nevertheless, by continuity,  the degenerate ground states
will
persist, and since the effects of 
flux loops and monopoles in the superconducting phase are local, 
the tunneling between them is still
given solely by the flux lines traversing the whole torus.
In a real superconductor with unit charge quasiparticles 
present, the situation is more complicated, and the ground state is 
determined by competing tunneling processes~\cite{hos2004}.

Allowing, for simplicity, tunneling only in the $2$-direction and
restricting the Hilbert space to the two ground states (which is
appropriate at low temperatures),
we may write down a tunneling Hamiltonian with
elements $H_{11} = H_{22} = E_0$, $H_{12} = H_{21} = -\Delta$
in the flux basis $\ket{f_1}$,
which
is thus diagonal in the eigenstates
$\ket{\pm} = (\ket{0} \pm \ket 1 )/\sqrt{2}$ of $\Omega_{10}$, with
eigenvalues $E_\pm = E_0 \mp \Delta$.  From the definition of the
twisted and untwisted partition function, it now follows,
\bee{split}
\bra\pm e^{-\beta H}\ket\pm = Z_{ev.} \pm Z_{odd} 
= e^{-\beta(E_0 \mp \Delta)}\, ,
\ee
which gives the final formula for the energy splitting,
\bee{twloop}
e^{-2\beta\Delta}= \frac {Z_{ev.} - Z_{odd}}  {Z_{ev.} + Z_{odd}}  
=\tilde W_\mu \, ,
\ee
where the last equality follows since the total flux operator
$\Phi_0 M_\mu$ takes the values $0$ and $\pi$ for $q=2$. 
Not only does this result allow the gap $\Delta$ to be explicitly
calculated from $\tilde W$.  It also implies, via \Eq{W-scaling}, a
scaling relation, $\Delta = \hat\Delta_\pm(L/\xi)/L$.
Expressed in the original variables, $\tilde W$ takes the form
$ \tilde W =\av{ P_1(0,x_2) P_1^\dagger(\beta,x_2) }$,
\emph{c.f.}\ \Fig{tunnel}~\footnote{
This operator differs from 1 only at the one
particular link where the $A$ field is twisted. This explains,
in terms of the original variables, why $\tilde W$ has no perimeter law.}.
The generalization to the full 4 state system is straightforward -- 
the splitting pattern is now $(-2\Delta, 0,0,2\Delta)$, 
while the relation \pref{twloop} between $\Delta$ and $\tilde W$
still holds.
It is also possible to generalize to higher $q>1$ and other topologies.

Deep into the superconducting phase we can estimate $\tilde W$ by assuming that
straight, and statistically independent, paths dominate the partition
sum.  An easy calculation using \pref{stronga} gives
$
  \Delta_0 = (L/a^2) e^{-c L/a}
$,
where $a$ is the lattice spacing, and $c$ is the line
tension of the flux lines ($c = \Phi_0^2/2g$ for $\lambda \ll a$ and
$J\pi \ln(\lambda/a)$ for $\lambda \gg a$).
To extend this analysis away from the small $\lambda$ and small
$g$  part  of the phase diagram, and especially to the region close
to the phase transition, we  match the splitting $\Delta_0$ deep in
the phase to the scaling expression given in \Eq{W-scaling},
leading to
\bee{scal}
  \Delta = \frac {L}{\xi^2} e^{-c' L/\xi}\, ,
\ee
for $\xi \lesssim L$, where $\xi \sim |\delta|^{-\nu}$ is now a physical
correlation length.
Thus the energy splitting gets renormalized and eventually closes as
the transition is approached.  The exponentially small splitting has
been predicted by Wen, whereas here we obtain also a nontrivial
prefactor.

\begin{figure}
\centerline{
\includegraphics[width=5cm]{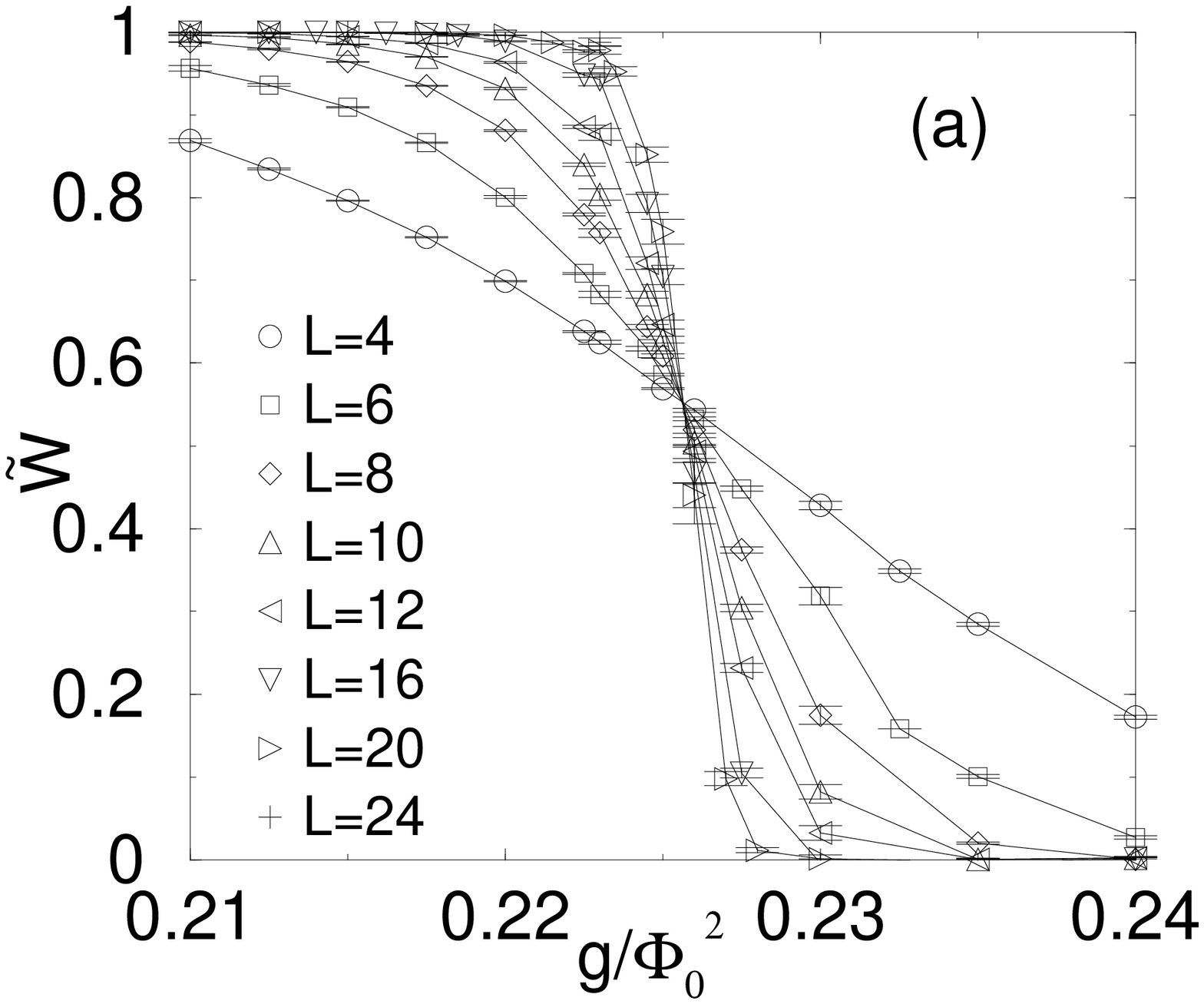}
\includegraphics[width=5cm]{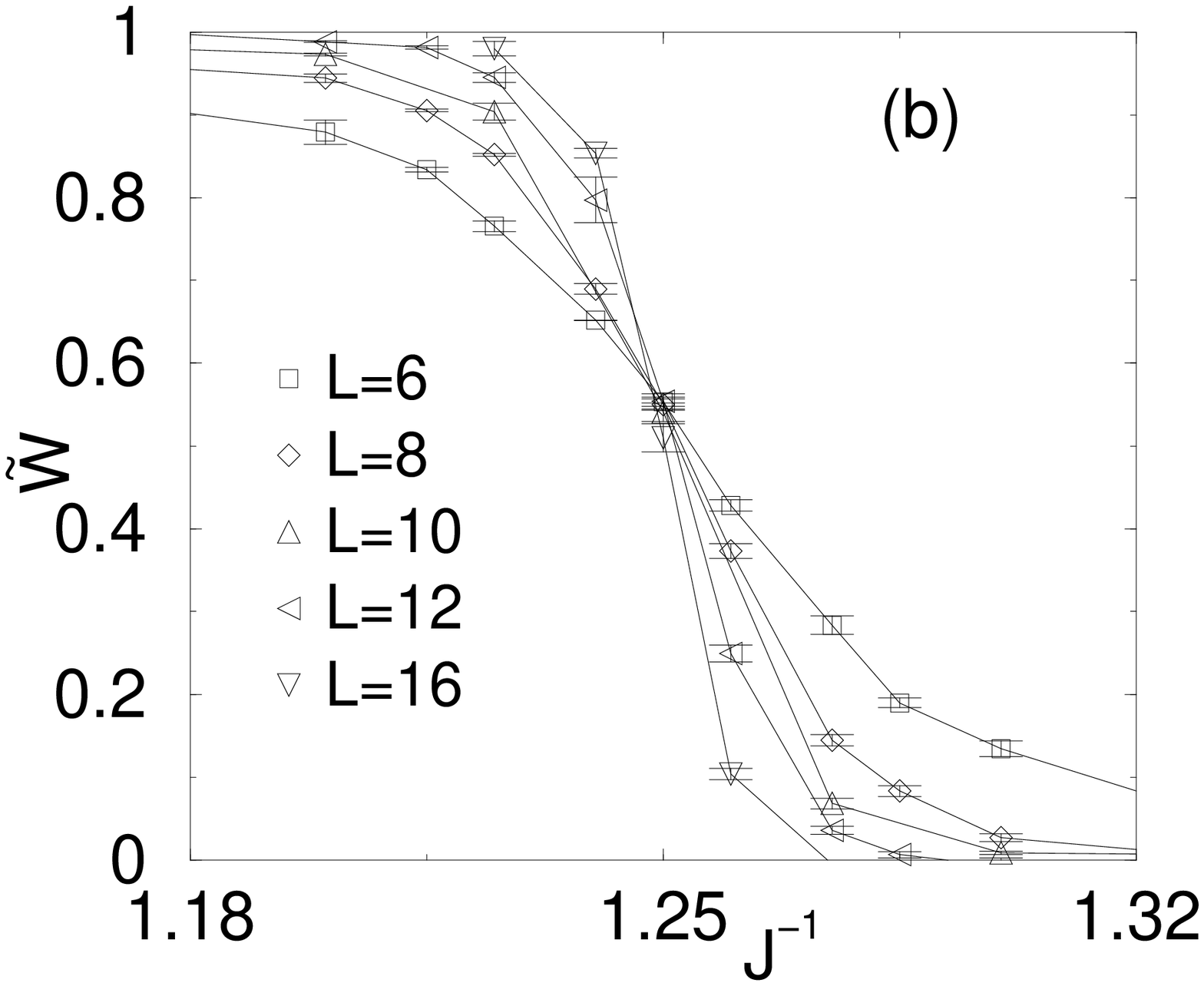}
}
\caption{
  \label{lambda0}
  $\tilde W$ as function of $g$ for the $\lambda \to 0$ case (a),
  and of $J^{-1}$ for $\lambda = 0.5$ (b).  The different curves cross
  right at the phase transition, where $\tilde W$ becomes independent
  of the system size, according to \Eq{W-scaling}.
}
\end{figure}

Equation~\Ref{twloop} enables us to explicitly calculate the splitting
 $\Delta$ numerically using Monte Carlo simulations.
 We
simulate the system in the flux line -- monopole representation given
by \Eq{duals} or \Ref{stronga}, using a variant of a recently
described worm cluster update Monte Carlo algorithm~\cite{worm},
properly adapted to long range interactions and the existence of
magnetic monopoles.  This algorithm naturally includes global moves
which change $M_\mu$ ({\it i.e.}, we allow twists in the boundary conditions
\Ref{tbc}) that are necessary for the calculation of $\tilde W$,
whereas in a conventional Metropolis algorithm the acceptance ratio
for such moves becomes exponentially small with increasing lattice
size.
For convenience we allow twists not only in the time direction but
also in space, {\it i.e.}, we simulate the system in a grand canonical
ensemble.  We have checked that this does not change our results.
The details of the numerical methods will be described
elsewhere~\cite{vester2}.

\begin{figure}
\centerline{\includegraphics[width=8cm]{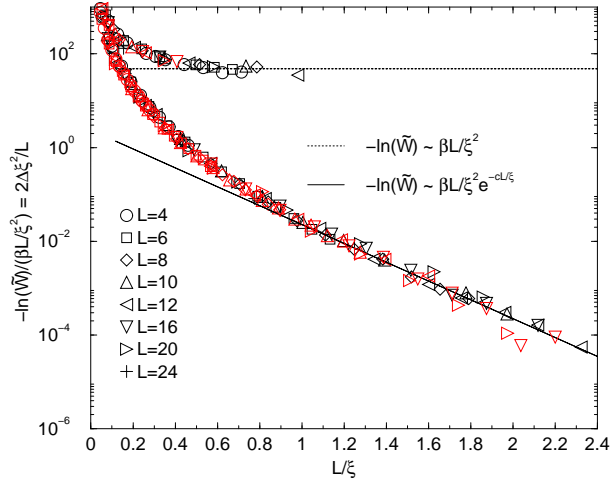}}
\caption{
  \label{GSsplittwist}
  Test of \Eq{scal} for $\lambda = 0$.  When plotted against $L/\xi$,
  the combination $2\Delta \xi^2/L$ falls onto two separate branches.
  The data for $g<g_c$ shows an exponential dependence (indicated by
  the solid straight line) in agreement with \Eq{scal}.  The data for
  $g>g_c$ shows an area law indicated by the horizontal dotted line.
  Black symbols are for $\beta=L$ and red (gray) symbols are for
  $\beta=2L$.  }
\end{figure}

\begin{figure}[b]
\centerline{\includegraphics[width=8cm]{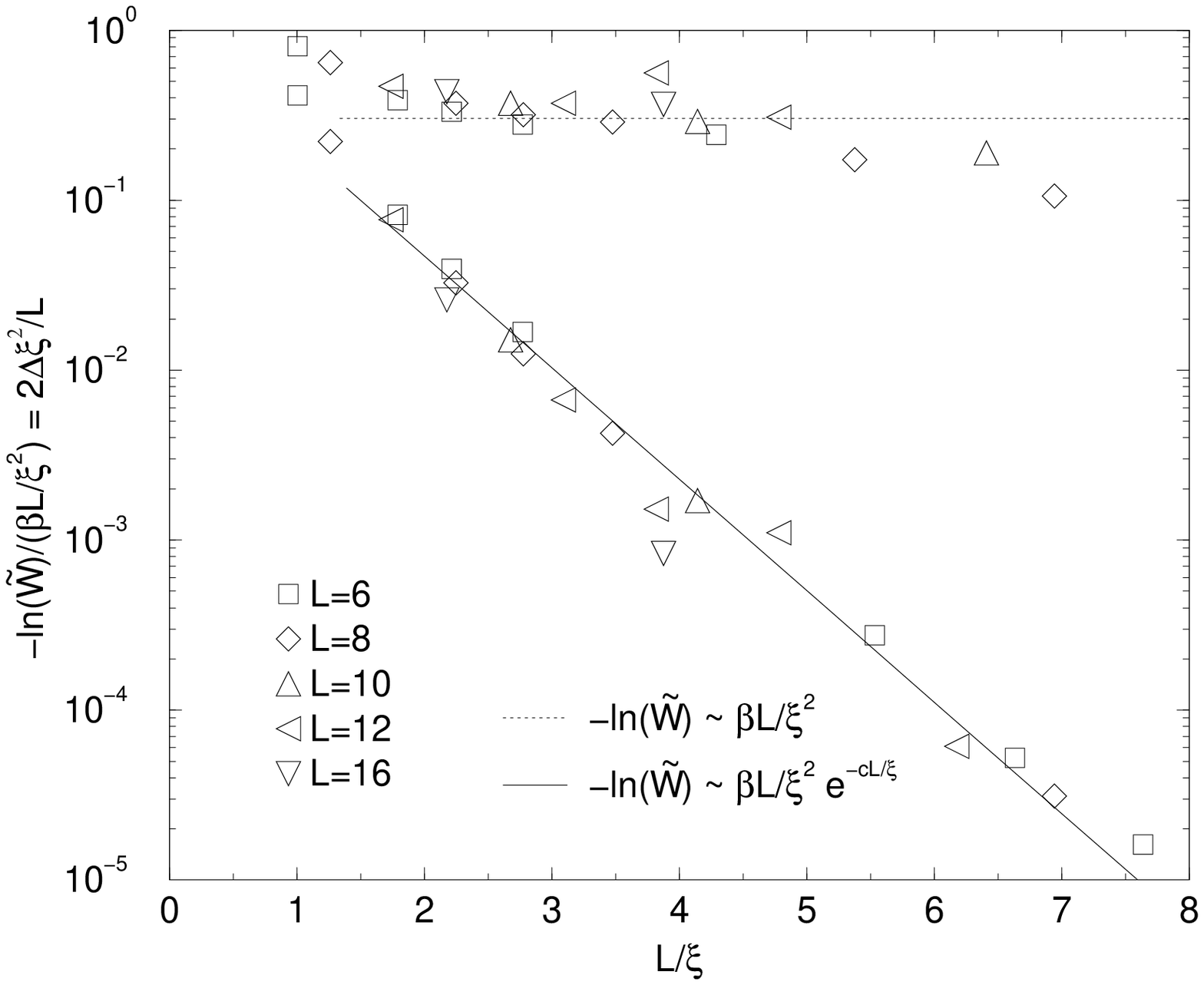}}
\caption{
  \label{finitelambda} 
  Same as \Fig{GSsplittwist}, but for $\lambda=0.5$.  }
\end{figure}

We have carried out simulations for constant $\lambda=0,0.5,1,2$,
varying $J$ in each simulation.
For a given $\lambda$, the critical value $J_c$ (or $g_c$ in case
$\lambda=0$) of the deconfinement transition is located using the
nonlocal order parameter $\tilde W$, as shown in \Fig{lambda0}.
According to \Eq{W-scaling}, $\tilde W$ for different system sizes
should cross right at the transition.
In the topologically ordered phase $\tilde W$ tends to one, implying
that $\Delta \to 0$ and the states becomes close to degenerate.  The
exponential dependence of $\Delta$ on $L$ in \Eq{scal} is obtained in
\Fig{GSsplittwist} for $\lambda = 0$ and in \Fig{finitelambda} for
$\lambda=0.5$.  We plot $2\Delta \xi^2/L$ vs $L/\xi$, where the
correlation length is given by $\xi = A |\delta|^{-\nu}$.  The
correlation length exponent $\nu$ is adjusted until the data collapse
onto two branches (for positive and negative $\delta$), which happens
for $\nu=0.63$, consistent with the 3D Ising critical behavior
expected for $q=2$.  The proportionality constant $A$ is fixed by
\Eq{scal}.  The straight line in the lin-log plot indeed demonstrates
the exponential scaling in the topologically ordered phase, whereas an
area law indicated by the horizontal line is obtained in the confined
phase.  Note that
a $e^{-cL^2/\xi^2}$ behavior characteristic of a spontaneously 
broken discrete symmetry is excluded. 
The results for $\lambda = 1$ and $2$ are very similar.
In principle, the simulations done for $\beta = L$ cannot rule out a
weak dependence on $\beta/L$ in \Eq{scal}.  Therefore we have also
calculated $\Delta$ from anisotropic systems with $\beta=2L$, see
\Fig{GSsplittwist}, showing that any such residual dependence on
$\beta/L$ must be extremely weak.

In summary, we have shown that the ground state energy splitting
$\Delta$, between the almost degenerate ground states in the
topologically ordered phase of a lattice superconductor is 
simply related to  a novel nonlocal order parameter $\tilde W$ of the
deconfining transition. Using a Monte Carlo algorithm 
we calculated $\tilde W$, demonstrated that it is a good order parameter
for the phase transition, and established that $\Delta$ has 
an exponential dependence on system size in full agreement with the
vortex tunneling mechanism proposed by Wen.
This suggests that the lattice superconductor shows the characteristics
of topological order all the way to the phase transition.  Let us
finally remark that the methods used in this paper should be
applicable also to a model with quenched impurities.

\acknowledgments

We thank Asle Sudb{\o} and Shivaji Sondhi for valuable discussions
and comments on the manuscript.
Support from the Swedish Research Council (VR) and the G{\"o}ran
Gustafsson foundation is gratefully acknowledged.


\newcommand{\prb}{Phys.\ Rev.\ B}
\newcommand{\prd}{Phys.\ Rev.\ D}
\newcommand{\pre}{Phys.\ Rev.\ E}
\newcommand{\prl}{Phys.\ Rev.\ Lett.\ }
\newcommand{\rmp}{Rev.\ Mod.\ Phys.\ }

\bibliographystyle{unsrt}

\bibliography{compact}

\end{document}